# Optimizing Electric Taxi Charging System: A Data-Driven Approach from Transport Energy Supply Chain Perspective

Yinghao Jia
Department of Industrial Engineering
Tsinghua University
Beijing, China

Yide Zhao, Ziyang Guo, Yu Xin
Smart Cities Research Group
Sparkzone Institute
Beijing, China
(research@sparkzone.org)

Huimiao Chen
Department of Electrical Engineering
Tsinghua University
Beijing, China

*Abstract*— **In the last decade, the development of electric taxis has motivated rapidly growing research interest in efficiently allocating electric charging stations in the academic literature. To address the driving pattern of electric taxis, we introduce the perspective of transport energy supply chain to capture the charging demand and to transform the charging station allocation problem to a location problem. Based on the P-median and the Min-max models, we developed a data-driven method to evaluate the system efficiency and service quality. We also conduct a case study using GPS trajectory data in Beijing, where various location strategies are evaluated from perspectives of system efficiency and service quality. Also, situations with and without congestion are comparatively evaluated.**

*Keywords--* **Electric taxis; transport energy supply chain; charging station planning; data-driven approach.**

## I. Introduction

Compared with conventional gasoline-powered taxis (CTs), environmentally friendly electric taxis (ETs) gains their popularity due to lower pollution emission and energy consumption. Recently, governments around the world are encouraging the development of ETs by providing subsidies and establishing more charging stations. However, finding the optimal allocation of charging stations so as to maximize the traffic system's efficiency is still an open question.

In the last decade, the development of ETs has motivated a rapidly growing research interest in efficiently allocating electric charging stations in the academic literature. Most, if not all, of existing studies mainly concentrate on accurately capturing ETs' charging behavior. Thus, several model-based and data-based methods are proposed to solve this allocation problem. In model-based approaches, flow-capturing models [1]-[3] and network-equilibrium models [4]-[6] are often adopted. However, these models introduce too many assumptions of ET drivers' behavior; for example, they constrain the place and frequency of charging, which make these models inconsistent with the situation in the real world due to the complexity and the uncertainty of ET drivers' choice.

Based on the Global Positioning System (GPS) trajectory data, researchers turn to data-based methods for a better understanding of ET drivers' behavior. These studies mainly utilize vehicle trajectory data extracted from conventional taxi drivers and the key assumption in their methods is that the driving patterns of CTs and ETs are identical, which means ETs are recognized as electrified CTs [7]-[10]. However, this assumption may not hold for some situations. References [11] and [12] compare the GPS trajectories of CTs and ETs. The authors demonstrate that the driving patterns of these two kinds of taxis are significantly different because the battery capacity of a typical ET is limited and the drivers may suffer from "range anxiety", that is, the ET drivers need to decide whether to deliver service of a certain order if the expected traveling distance of this order approximately reaches the limit of battery remaining capacity. On the other side, ET drivers need to get their vehicles recharged when the battery capacity is low, which indicates that ETs need to periodically go to electric charging stations. Since the range limit of a typical ET battery is relatively low, i.e., 150 kilometers, such charging demand of ETs can occur once or twice a day. Differently, drivers of CTs can meet customers' demand with less range anxiety because there are enough gasoline stations in the city and the range of a typical CT can reach 500 kilometers. Although References [11] and [12] provide baseline and give inspirations for our research that ETs and CTs have different driving patterns, they only focus on data analysis of electric taxis instead of the solution of efficiently allocating charging stations.

To address the significantly different driving patterns of CTs and ETs, in this paper, we introduce the idea of transport energy supply chain to capture ET drivers' driving pattern and their charging demand. In our approach, charging stations are regarded as energy suppliers and transport needs of customers are regarded as energy demanders. Moreover, we connect individual trips of a vehicle to a trip-chain so as to capture the charging demand. Based on trip chains, we can extract the charging demand of vehicles once they are electrified from real GPS trajectory data, which makes it possible to quantitatively valuate the effect of the charging stations allocation. As a result, we reduce the difficulty of capturing ETs' driving pattern because the potential but more essential factors—supply and demand—are considered and ETs serve as a medium of transporting energy from charging station to customers. Using the transport energy supply chain, we further transform the charging station allocation problem to a location problem covering all possible ET service areas, which can be efficiently solved using existing built-in methods provided by optimization software packages. Besides, to evaluate the system efficiency and service quality, we model this location problem using two models in Reference [13]: (1) P-median model to minimize the demand-weighted total distance; and (2) Min-max model to constrain the maximum distance between each demand node and charging station to a reasonable range. Based on the two models, we propose a data-driven method to properly allocate the electric charging stations, which is tested in a case study using 39,053 GPS records of urban CTs in Beijing, China. We also take into consideration the possible influences of congestion and the number of charging stations in our sensitivity analysis.

The main procedures and contributions of this research are concluded as follows:

---We introduce the transport energy supply chain to capture the driving patterns of ETs, which significantly reduces the complexity of directly analyzing every individual ET trip;

---Based on trip chains, we proposed a method to estimate the charging demand of ETs and transform the charging station allocation problem into a location problem, which are modelled using P-median and Min-max models;

---We conduct a numerical case study to evaluate the system efficiency and service quality of our models if CTs are electrified. In our case study, we extract the charging demand based on 39,053 GPS records of urban CTs in Beijing, China;

---We analyze the impact of congestion and the number of total charging stations that are built. And we notice that building more stations in the downtown may lift the system efficiency when congestion is taken into consideration of the P-median model. While the impact on the maximum distance from certain demand point to the station may be limit based on the results of Min-max model. Besides, the marginal benefit brought by building more charging stations gradual decreases, which may underutilize the usage of public resources.

The reminder of this paper is organized as follows. In section II, we introduce the transport energy supply chain and demonstrate the process of transforming the charging station allocation problem into a location problem. In Section III, we conduct a case study based on real data and use evaluation indexes to compare the P-median and the Min-max models. Finally, we conclude the discussion in Section IV.

## II. MODELLING APPROACH

As addressed in References [11] and [12], the driving patterns of ETs and CTs are significantly different. That is, drivers of ETs, with the anxiety of running out of electricity, can only deliver service within a certain range near the charging stations. On the other side, drivers of CTs can pick up any customer in the city because of enough gasoline stations and a large service range of CT which can reach as far as 500 kilometers. As a result, the assumption of unchanged driving patterns between CTs and ETs does not hold for most empirical studies.

However, directly analyzing the driving pattern of ETs is extremely difficult due to its uncertainty and randomness. We note that a driver of a typical full charged ET with the range of 150km can usually meet one customer's demand. That is, the range anxiety often works when the driver is going to reach the range limit. To implement this key observation, we, therefore, introduce the idea of transport energy supply chain, which focuses on the essential perspective of suppliers and demanders. In our models, ET, regarded as an important medium, serves the purpose of transporting energy from the suppliers, the electric charging stations, to the demanders, customers' transport needs.

In the transport energy supply chain, we group individual trips into many trip-chains using the following method: For a certain fully charged ET, we connect the destination of its first trip to the origin of the second trip, forming a continuous chain. Then we continue to connect the third trip to the current trip chain if the following conditions are simultaneously met: (1) The starting service time of the third trip is close to the end time of the second trip, i.e., within 5 minutes; (2) The origin of the third trip is near to the destination of the second trip, i.e., within similar longitudes and latitudes; and (3) The total travel distances of the current trip chain and the third trip are less than the maximum capacity of an ET's battery, i.e., 150 kilometers. Moreover, we repeat the procedure of connecting trips of this ET until the distance of this extending trip chain reaches 150 kilometers, and the connected chain is regarded as a demand for energy, that is, this ET needs to find the nearest charging station to recharge in order to sustain its service.

The biggest benefit of the transport energy supply chain is that we can capture the charging demand of ETs without analyzing the specific trajectories of every ET. As a result, we can have several charging demand caused by ET. Then the problem of optimizing the charging station allocation that covers the origin and destination of every trip chain can be transformed into a location problem. In this paper, we adopt two classical location models with different objectives to measure the quality of service: (1) P-median model to minimize the demand-weighted total distance; and (2) Min-max model to constrain the maximum distance between each demand node and charging station to a reasonable range. Therefore, we can introduce the following two indexes to evaluate the performance of certain charging station allocation plan based on the objectives of P-median and Min-max models: (1) Average distance from OD points to charging station, which reflects the efficiency of the whole system; and (2) The maximum distance from O/D points to charging stations, which reflects the quality of service of this system.

Firstly, we model this location problem using P-median model as follows.

$$\min \sum_{j \in J} \sum_{i \in J} h_i d_{ij} Y_{ij} \quad (1)$$

$$s.t. \sum_{j \in J} Y_{ij} = 1 \quad \forall i \in I \quad (2)$$

$$Y_{ij} - X_j \leq 0 \quad \forall i \in I, j \in J \quad (3)$$

$$\sum_{j \in J} X_j = m \quad (4)$$

$$X_j \in \{0,1\} \quad \forall j \in J \quad (5)$$

$$Y_{ij} \in \{0,1\} \quad \forall i \in I, j \in J \quad (6)$$

TABLE I. VARIABLES OF OUR MODELS

| Symbol | Meaning |
|---|---|
| $h_i$ | Represents the intensity of demand at location $i$ |
| $d_{ij}$ | Represents the equivalent distance between location $i$ and candidate charging station $j$ |
| $Y_{ij}$ | Represents an assignment variable. When it equals 1, orders at location $i$ is assigned to a charging station at location $j$, and 0 otherwise |
| $X_j$ | Represents a binary decision variable. When it equals 1, we locate a charging station at location $j$, and 0 otherwise |
| $m$ | Represents the number of charging stations that are built |
| $J$ | Represents the set of candidate charging stations in the studying area |
| $I$ | Represents the set of demand location in the studying area |

In the above model, the objective function minimizes the demand-weighted total distance with unbalanced demand. Constraints **(2)** stipulates that every demand node is assigned to a charging station, that is, we assume that every charging demand is met. Constraints **(3)** stipulates that a charging assignment can be assigned to the location where there is a charging station. Constraints **(4)** limit the total number of charging stations that can be built in the studying regions. Finally, Constraints **(5)** and **(6)** restrict the decision variables to be binary. As a result, there are $|J|^2+|J|$ decision variables and $|J|^2+|J|+1$ constraints in the P-median model.

Secondly, we introduce the Min-max model to constrain the maximum distance between each demand node and charging station to a reasonable range.

$$\min \max \{d_{ij} Y_{ij}, \forall i, j \in J\} \quad (7)$$

s.t. **(2)- (6)** ,

where the definitions of variables are the same as the P-median model. Based on these two models, we can measure efficiency and quality of service among given charging station allocation plans.

Once we capture the charging demand of vehicles, we can use standard built-in methods provided by binary integer programming software packages to obtain a high-quality allocation of charging stations. In fact, according to our experience, existing methods can solve the P-median and the Min-max models efficiently within one hour for all the instances that we considered.

III. CASE STUDY

*A. Raw data analysis*

With the rapid development of information technology, researchers now can take the advantages brought by data-driven methods to understand customer origin-destination (OD) pairs based on mass GPS data. For example, Reference [10] develops an approach to capture customers' demand through spatially clustering GPS data.

In our case studies, we characterize customers' demand based on real world data collected from 39,053 taxis in Beijing, China. These data are collected through mobile internet applications and onboard device in taxis within the fifth ring of Beijing——with latitude between 39.75N and 40.03N and longitude between 116.2E and 116.55E—— in between the time period of May 4th 2016 and May 31st 2016. For each individual record in our dataset, it collected the information of the OD pair with starting and ending service time, the longitudes and latitudes of origin and destination, and the travel distance. A sample of our data is represented in Table 1. It means the driver spends around 20 minutes delivering a customer from suburb called Anding Town to a place near Beijing Nanyuan Airport on May 4th, 2016 with a travel distance of 22.365 kilometers.

TABLE II. RECORD SAMPLE

| O Time | Longitude | Latitude | D Time | Longitude | Latitude | distance |
|---|---|---|---|---|---|---|
| 20160504 08:45:35 | 116.4915 | 39.6175 | 20160504 09:03:07 | 116.4331 | 39.8042 | 22.365 |

From more than 23 million OD records in our dataset, we further filter our useable data through modifying the following types of data: (1) Missing data. We apply method of interpolation to supplement missing points in the trajectories, but directly delete the missing data if the data collectors fail to send GPS information to the cloud in more than 5 minutes; (2) Drafting data. We utilize GPS mapping technology to correct drifting points using digital maps and geographic information system (GIS) system to limit the error range of GPS points; and (3) Duplicated data. We delete occasion duplicated data caused by the information collecting system.

After the above data-processing, we then extract 3,000 trip chains with each chain approximately 150 kilometers but strictly less than 150 kilometers in our case study. In this way, each trip-chain represents a charging demand from certain vehicle. Moreover, the destinations of all trip-chains form the set of demand points, *I*. Theoretically, all demand points can be a candidate for possible charging station, but the aim of building a charging station is to meet the demand in certain region characterized by distance. So we introduce K-means to cluster all those demand points into a smaller subset distributed in the city, which forms the set of possible locations for charging stations, *J*. Then the distance from points in *I* to points in *J* can be calculated through Manhattan distance. Moreover, we notice that the intensity demand of certain location is almost the same with deviation less than 10 percentages. Therefore, in our case study, we further set $h_i$ to one in the P-median model. Based on these discussion, we can finally calculate the values of parameters in the P-median and the Min-max models. Then all that remains is to solve the two models taking into consideration some important factors that will significantly influence the quality of service of a certain charging station allocation plan. In the following sections, we try to analyze the impact of congestion and the number of total charging stations that are built.

*B. Comparision between models*

In our case study with 3,000 trip chains, the size of the set of demand points is 6,000. And we choose 100 candidate charging stations after implementing the K-means method. We further assume that the government of Beijing is going to build 30 charging stations within the firth ring, that is, *m* is set to 30 in this simulation. Then we use CPLEX [14] as the solver to solve the P-median and the Min-max model described in Section II.

In this simulation, all trip-chains are extracted using data collected during non-congestion period, that is, the connected trips are collected during 8:00am to 6:00pm and 8:00pm to 6:00am. The results of the two models are presented in Figure 1. In Figure 1, a star and a red dot represent the location of a charging station location selected by the P-median and the Min-max models, respectively. As we can see from Figure 1, the red dots scatter around the city, while the stars are more intensive in the downtown. This observation implies that the tradeoff between the system efficiency and the quality of service should be taken into consideration for decision making.

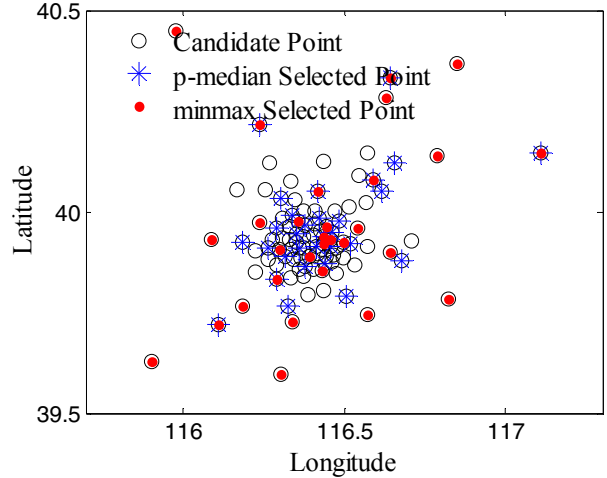

Fig. 1. The charging station allocation plans of the P-median and the Min-max models.

We then try to analyze the impact of the number of total charging stations. To achieve this, we consider increase the value of *m* from 30 to 60. Then we solve the corresponding P-median and Min-max models to see how the number of charging stations may influence the average distance from the demand points to the charging stations. In Figure 2, the vertical axis represents the average distance from the demand points to the charging stations based on the results of P-median models, but with different values of *m*. As we can see from Figure 2, as the value of *m* increases, the average distance decreases. This observation is reasonable because more charging station means drivers of ETs can find a near charging station more easily when they need to recharge their taxis. However, we also note that the speed of average distance decreasing gradually declines.

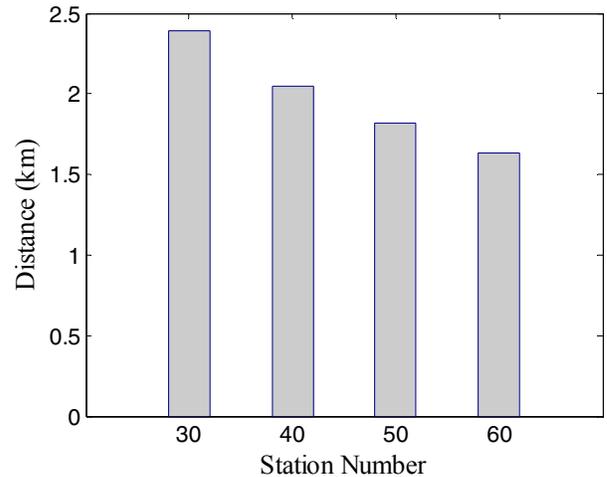

Fig. 2. The impact of increasing charging stations on the average distances using the results of P-median models.

On the other side, we also test the impact of increasing charging stations on the maximum distance from demand points to the charging stations using the results of Min-max models. And Figure 3 represents the results, where the maximum distance decreases gradually as the number of stations increases. However, the speed of maximum distance decreasing gradually declines. Both of the observations in Figures 2 and 3 implies that the marginal benefit on lifting the system efficiency and quality of service, brought by building more situations will gradually decrease. That is, directly building plenty of charging stations may underutilize the usage of public resources.

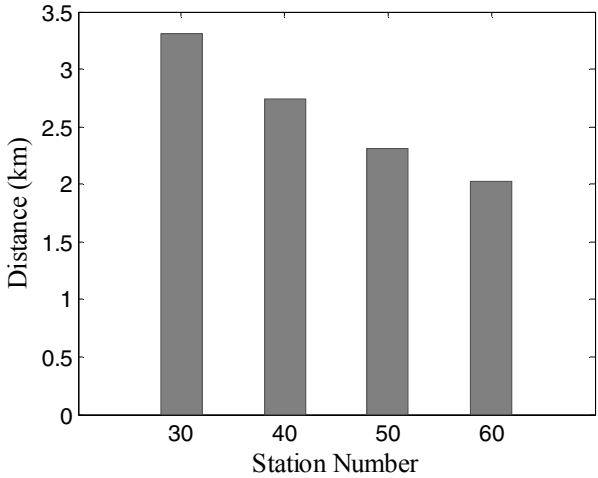

Fig. 3. The impact of increasing charging stations on the maximum distances using the results of Min-max models.

Here we utilize congestion factor $\sigma$ to measure traffic congestion's influence, i.e. the delay effect caused by traffic congestion usually happening at rush hour in downtown area, leading to the same distance traveling much more time and consuming more energy.

TABLE III. CONGESTION FACTOR DESCRIPTION

| Symbol | Meaning |
| --- | --- |
| $\sigma=1.5$ | Trips happen within the 3rd ring and 7:00am-9:00am & 18:00pm-20:00pm |
| $\sigma=1.2$ | Trips happen between the 3rd ring and the 4th ring within 7:00am-9:00am & 18:00pm-20:00pm |
| $\sigma=1$ | Other trips |

When congestion is taken into consideration of the P-median model, we notice that more stations are located in the downtown, as shown in Figure 4, while the impact on the maximum distance from certain demand point to the station can still be limited based on the results of Min-max model. Thus, taking congestion into consideration provides a more realistic solution, but, more importantly, increases system efficiency without diminishing service quality.

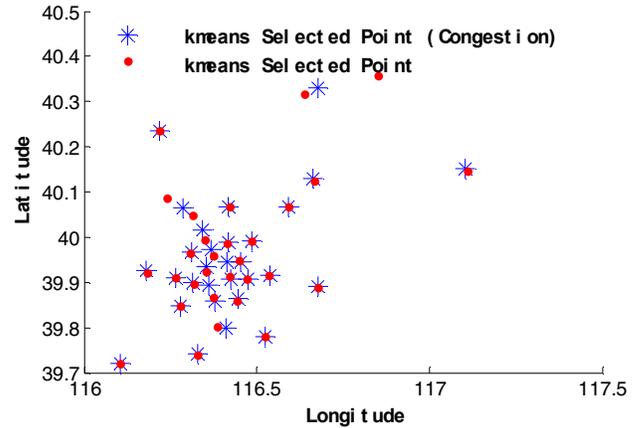

Fig. 4. The impact of congestion on charging stations allocation

IV. CONSLUSIONS

In this paper, we try to find an efficient electric charging station allocation plan so as to meet the rapidly growing demand of ETs in the city. To address the significant different driving patterns of CTs and ETs, we adopt the perspective of the transport energy supply chain that charging stations serve as energy suppliers and customers serve as energy demanders. In this way, we can easily capture ETs' charging demand without directly analyze the individual trajectories of ETs. Based on trip-chains, we further transform the original allocation problem into a location problem that covers all possible demand points in the studying areas. To evaluate the performance of certain charging station, we introduce the P-median and the Min-max model to allocate charging stations to provide efficient and quality service. Based on the two models, we propose a data-driven method to properly allocate the electric charging stations, which is tested in a case study using 39,053 GPS records of urban CTs in Beijing, China. We also take into consideration the possible influences of congestions and the number of charging stations in our sensitivity analysis.

V. ACKNOWLEDGEMENT

The authors appreciate the help offered by Sparkzone Institute and the data supported by Tsinghua Daimler Sustainable Transportation Laboratory.